\documentclass[ twocolumn,08 pt, amsmath,amssymb, aps,fleqn]{revtex4-1}
\usepackage[hidelinks]{hyperref}
\usepackage[english]{babel}
\usepackage{mathtools}
\usepackage{tabularx}
\usepackage{multirow}
\usepackage{comment}
\usepackage{graphicx}
\usepackage{color}
\usepackage{soul}
\begin{document}
\preprint{APS/123-QED}
\title{Ponderous impurities in a Luttinger liquid}
\author{Joy Prakash Das}   \author{Girish S. Setlur}\email{gsetlur@iitg.ernet.in}
\affiliation{Department of Physics \\ Indian Institute of Technology  Guwahati \\ Guwahati, Assam 781039, India}

\begin{abstract}
In this work, analytical expressions for the Green function of a Luttinger liquid are derived with one and two mobile impurities (heavy particles) using a combination of bosonization and perturbative approaches. The calculations are done in the random phase approximation (RPA) limit using the powerful non-chiral bosonization technique (NCBT) which is nothing but the resummation of the {\it{most singular parts}} of the RPA terms of the Green function expanded out in powers of the forward scattering between fermions {\it{with the source of inhomogeneities treated exactly}}. The force acting on the heavy particle(s) is studied as a function of its terminal velocity, both in the linear and non-linear regime.
Linear mobility (which is valid for impurities moving much slower than a certain cross-over speed) has a power-law temperature dependence whose exponent has a closed algebraic expression in terms of the various parameters in the problem. This expression interpolates between the ballistic regime of no-coupling with the fermions and the no-tunneling regime. When the speed of the impurity is much larger than this cross-over speed, the applied force depends non linearly on the speed and this too is a power-law with a closely related exponent. The case of two mobile impurities is also studied whose mobility exhibits peculiar resonances when their mutual separation is appropriately chosen.
\end{abstract}

\maketitle
\section{Introduction}

One of the major themes in condensed matter physics is the study of the effect of impurities in  quantum systems. Various types of impurities are studied in the existing literature : Coulombic impurities \cite{brum1984screened} and Kondo impurities \cite{nagaoka2002temperature, ujsaghy2000theory} to name a few. Impurities in solid state systems are commonly immobile and are considered as static perturbations. On the other hand, mobile impurities are more typically studied in fluidic systems \cite{baym1967phonon}. The well-known example in this regard would be the motion of a heavy particle in a three dimensional quantum fluid \cite{prokof1995effective, prokof1993diffusion}. The presence of an impurity can bring drastic changes to  interacting systems, especially if the systems are one dimensional \cite{kane1992transport}, whose physics is quite different from higher dimensional systems, better described by the Luttinger liquid model \cite{haldane1981luttinger}. With the advancement in micro/nano-fabrication \cite{brodie2013physics},  which has made the realization of 1D systems practical, there has been a growing interest towards problems concerned with the Tomonaga Luttinger (TL) liquids, the problem of an impurity in a TL liquid being one of them. The transport properties of TL liquids with localized impurities have been rigorously studied using bosonization, renornmalization group and other theoretical tools \cite{kane1992transport, kane1992transmission, furusaki1993single}.

More recently, the study of mobile impurities in one dimensional quantum liquids has been an active area of research\cite{neto1996dynamics, caldeira1995motion,  tsukamoto1998critical, fukuhara2013quantum,astrakharchik2004motion, mathy2012quantum}. Castro-Neto and Fisher \cite{neto1996dynamics} studied the dynamics of a heavy particle in a Luttinger liquid with repulsive interactions, besides analyzing the temperature dependence of the mobility of the particle. Caldeira and Castro-Neto computed the damping constant of a heavy particle, coupled to fermionic and bosonic environments in 1D \cite{caldeira1995motion}. Tsukamoto et al. obtained the exact critical exponents of various correlation functions of a TL liquid with a mobile impurity as functions of the impurity mass and momentum \cite{tsukamoto1998critical}. 
 Fukuhara et al. studied the quantum dynamics of a spin impurity as it propagates in a 1D lattice \cite{fukuhara2013quantum}. Astrakharchik and Pitaevskii \cite{astrakharchik2004motion}  predicted a power-law dependence of the drag force on a moving heavy impurity within Luttinger liquid on its velocity for small velocities. Girardeau and Minguzzi studied the problem of a moving impurity of finite mass  in a 1D gas of hardcore bosons, also known as the TG (Tonks-Girardeau) gas\cite{girardeau2009motion}. In the work done by Mathy et al. \cite{mathy2012quantum}, a phenomenon called quantum flutter was described for an impurity injected with finite momentum into a 1D quantum liquid. Schecter et al. realized that a constant force acting on an impurity in a 1D liquid leads to Bloch oscillations of the impurity around a fixed point, followed by energy release in the form of phonons \cite{schecter2012dynamics}.
 Lychkovskiy studied mobile impurities in a 1D quantum fluid at zero temperature and concluded that the velocity of the impurity at infinite time does not vanish at zero temperature, which is not the case at finite temperature \cite{lychkovskiy2014perpetual,lychkovskiy2015perpetual}.

 In this work, the Green functions of a Luttinger liquid in presence of a slowly moving heavy particle (or a pair of them) is calculated using a combination of perturbative approaches and  the non-chiral bosonization method \cite{das2016quantum,das2017one}. Using the Green function, the force acting on the heavy particle is calculated in terms of its terminal velocity, both in the linear and non-linear regime.  Mobility is calculated for the linear regime, shedding light on temperature dependence and  mutual interaction strength between the fermions. Our results are qualitatively consistent with the highly cited work on the subject \cite{neto1996dynamics} both at low and high temperatures. Our findings are also consistent with the impenetrable impurity moving ballistically (mobility diverges) at low temperatures so long as the fermions are mutually repelling. At higher temperatures small compared to the Fermi energy the mobility saturates to a constant value.

\section{System description and method of solution}
Consider a Luttinger liquid in one dimension with forward scattering short range mutual interactions \cite{giamarchi2004quantum} in the presence of a heavy particle (impurity) moving with speed small compared with the Fermi velocity $ v_F $. The full  generic-Hamiltonian of the system(s) under study (before approaching the RPA limit) is (are),

\begin{equation}
\begin{aligned}
H =& \int^{\infty}_{-\infty} dx \mbox{    } \psi^{\dagger}(x) \left( - \frac{1}{2m} \partial_x^2 + V_0\delta(x-X(t)) \right) \psi(x)\\
  &  + \frac{1}{2} \int^{ \infty}_{-\infty} dx \int^{\infty}_{-\infty} dx^{'} \mbox{  }v(x-x^{'}) \mbox{   }
 \rho(x) \rho(x^{'})
\end{aligned}
\end{equation}

where $V_0\delta(x-X(t))$ is the potential due to the impurity at position X(t) and $ v(x-x^{'}) = \frac{1}{L} \sum_{q} e^{ -i q(x-x^{'}) } v_q $ (where $ v_q = 0 $ if $ |q| > \Lambda $ for some fixed $ \Lambda \ll k_F $ and $ v_q = v_0 $ is a constant, otherwise) is the forward scattering mutual interaction. It is necessary to confine the study to the so-called RPA limit which means, among other things, working in the limit where the Fermi momentum and the mass of the fermion diverge in such a way that their ratio is finite (i.e. $ k_F, m \rightarrow \infty $ but $ k_F/m = v_F < \infty  $: units that make $ \hbar = 1 $, so that $ k_F $ is both the Fermi momentum as well as a wavenumber, are used). This amounts to linearizing the energy momentum dispersion near the Fermi surface and thereby leading to a feasible analytical solution.

The obvious method for studying this system is to observe that a mobile impurity moving with a speed much lower than the Fermi velocity may be regarded as being stationary to the lowest order approximation. The fermion Green function (with or without mutual interactions between fermions - in the former case, using  the non-chiral bosonization  \cite{das2016quantum,das2017one}) is computed with this assumption. In order to incorporate the effects of the non-zero speed of the impurity, an iteration of the relevant equations is performed wherein the zeroth order stationary Green functions are employed in order to compute leading corrections due to the finite speed of the impurity. The general Dyson's equation for the full Green function denoted by $ G(x,x';t,t') \equiv <T\mbox{  }\psi(x,t) \psi^{\dagger}(x',t')>  $ in terms of its counterpart $ G_{SCh}(x,x';t,t')  $ that assumes the impurity(s) is(are) stationary at(near) the origin is (here $C$ is the Keldysh contour),
\begin{equation}
\begin{aligned}
G(x,x';t&,t') = G_{SCh}(x,x';t-t') +  V_0  \int_{C} dt^{''} \mbox{  }\\
&\times\Big(G_{SCh}(x,X(t^{''});t-t^{''})G(X(t^{''}),x';t^{''},t') \\
&\hspace{1 cm}- G_{SCh}(x,0;t-t^{''})G(0,x';t^{''},t')\Big)\\
\end{aligned}
\end{equation}
Keeping in mind that the speed of the mobile impurity is small compared to the Fermi velocity, the leading approximation to the above full Green function would be,
\begin{equation}
\begin{aligned}
G(x,&x';t,t') \approx G_{SCh}(x,x';t-t') +  V_0  \int_{C} dt^{''} \mbox{  }\\
&\times\Big(G_{SCh}(x,X(t^{''});t-t^{''})G_{SCh}(X(t^{''}),x';t^{''},t') \\
&\hspace{1 cm}- G_{SCh}(x,0;t-t^{''})G_{SCh}(0,x';t^{''},t')\Big)
\label{Keldysh}
\end{aligned}
\end{equation}
Having obtained this, the force acting on the impurity may be computed as follows,
\begin{equation*}
\begin{aligned}
F_X =&- V_0 \left(\frac{d}{dx}\right)_{x=X(t)}<\rho(x,t)> \\
=& V_0  \left(\frac{d}{dx}\right)_{x=X(t)}G_{<}(x,x;t,t)
\end{aligned}
\end{equation*}
where $ G_{>} $ and $ G_{<} $ are the advanced and retarded Green functions respectively  and   $ G(x,x',t,t' ) = \theta(t-t')\langle\psi(x,t)\psi^{\dagger}(x',t')\rangle - \theta(t'-t)\langle\psi^{\dagger}(x',t')\psi(x,t)\rangle    $. The  RPA Green function
with and without mutual forward scattering interactions between fermions with stationary impurities has been
computed in a couple of recent works using the NCBT \cite{das2016quantum,das2017one}.

\subsection{Non chiral bosonization technique}
As in conventional bosonization schemes, the fermionic field operator is expressed in terms of currents and densities, although the field operator is now modified to include the effect of back-scattering by impurities.
\begin{equation}
\begin{aligned}
\psi_{\nu}(x,\sigma,t) \sim e^{ i \theta_{\nu}(x,\sigma,t) + 2 \pi i \lambda \nu  \int^{x}_{sgn(x)\infty}\mbox{ } \rho_s(-y,\sigma,t) dy}
\label{PSINU}
\end{aligned}
\end{equation}
Here $\theta_{\nu}$ is the local phase which is a function of the currents and densities ($\rho_s$) which is also present in the standard bosonization schemes \cite{giamarchi2004quantum}, which typically works for homogeneous systems. The new addition is the $\rho_s(-y)$ term which ensures the necessary trivial exponents for the single particle Green functions for a system of otherwise free fermions with impurities. Here $\nu$ can be 1 or -1 depending on right movers or left movers and $\lambda$ is either 0 or 1. Thus the standard bosonization scheme can be obtained by setting $\lambda = 0$. The factor $2 \pi i$ ensures that the fermion commutation rules are preserved. This field operator (annihilation), to be treated as a mnemonic to obtain the Green functions and not as an operator identity, is clubbed together with another field operator (creation) to obtain the two-point functions, the details being described in an earlier work \cite{das2016quantum}. 

When mutual interactions between fermions are absent, it is well known that the mobility is infinite (ballistic motion) for a homogeneous system. Upon introducing an impurity, the mobility gradually diminishes with increasing strength of the impurity and saturates to minimum non-zero value when tunneling across the impurity is forbidden. The NCBT is able to interpolate between these two extremes. The interesting question is how these expectations are modified upon inclusion of mutual interactions between fermions. The finding is that the mobility is a power law in the temperature with an analytically computable characteristic exponent so long as the speed of the impurity is much smaller than a certain cross-over speed. Alternatively, for a given applied force, the force is proportional to the terminal speed only when the temperature is much larger than a certain cross-over temperature (whose value is fixed by the applied force). For temperatures much lower than this scale, the linear mobility concept is no longer valid. All these assertions will be explicitly demonstrated in the discussion that follows.

\section{ Analysis and computations }

The analytical expressions for the Green functions of a Luttinger liquid in presence of one and two scalar impurities obtained in the cited central work \cite{das2016quantum} of the present authors is heavily relied upon in the present work. The technique of non-chiral bosonization used in this work provides an analytical expression for the most singular part of the RPA Green function of such systems which also happens to exhibit power law behavior analogous to homogeneous Luttinger liquids. 

 In the RPA limit, the assertion is: $ X(t) \rightarrow 0 $ even as $ k_F \rightarrow \infty $ so that $ |k_F X(t) | < \infty $. Hence quantities such as $ e^{ i k_F X(t) } $ that appear repeatedly in the calculations now make sense. The ansatz that amounts to asserting that the impurity executes a simple harmonic motion about the origin $ X(t) = \frac{v_X}{\omega} sin( \omega t) $ is used where $v_X $ is the maximum drift (steady state) velocity of the particle in response to an weak external force that is applied on it. The frequency $ \omega $ could represent one of two things -  a) if the applied force is sinusoidal in time, it is the frequency of the applied force b) if the force is independent of time, then $ |\hbar \omega| \sim T $ which is the temperature of the system. When the particle is moving very slowly i.e. $ |k_F X(t) | \ll 1 $ in other words
  $ \frac{k_F v_X}{T} \ll 1 $, calculations are done using the approximation $ e^{ i k_F X(t^{''}) } \approx 1 + i k_F X(t^{''}) $ which is inserted into the right hand side of eq.~(\ref{Keldysh}). This shows that the terminal speed is proportional to the applied force and the coefficient of proportionality is a power law in the dominant energy scale in this problem viz. temperature T, in this case.
   
   When the particle is not moving slowly i.e. $ \frac{k_F v_X}{T} \gg 1  $, it is not possible to expand in this fashion, but since $ e^{ i k_F X(t^{''}) } =  e^{ i  \frac{k_F v_X}{T} sin( T t^{''}) } $ where $ T $ is temperature in frequency units, this term rapidly oscillates and averages out to zero unless $ t^{''} \ll \frac{2\pi}{T} $. In this case we may write,  $ e^{ i k_F X(t^{''}) } \sim  e^{ i  k_F v_X  t^{''} } $ and rescaling $ t^{''} $ while performing the integral means the integrals in eq.~(\ref{Keldysh}) are going to be a power-law in the dominant energy scale in the problem which is now $ k_F v_X $ rather than temperature. This is the reason for the nonlinear power-law dependence of the force on the terminal velocity when $ k_F v_X \gg T $.
  
For a single mobile impurity, the externally applied force $ F_X $ on the impurity may be related to the drift velocity $ v_X $ quite easily. When fermions are not mutually interacting with one another this relation is 
\begin{equation}
F_X =     \frac{2V_0^2   k_F^2  }{\pi \left(V_0^2+v_F^2\right) } v_X\mbox{  }
\label{force1}
\end{equation}
where $ V_0 $ is the strength of the coupling between the heavy particle and the fermions.
Mobility $\mu$ is defined as the ratio between the terminal velocity of the impurity and the force acting on it so that,
\begin{equation}
\mu=     \frac{\pi \left(V_0^2+v_F^2\right) }{2V_0^2   k_F^2  }
\label{mobility1}
\end{equation}
The mobility diverges when the coupling between the impurity and the fermions vanishes, implying ballistic motion of the impurity i.e. the impurity accelerates in response to the external force rather than reaching a terminal velocity. Conversely, when the coupling diverges - which means no tunneling of fermions through the impurity is permitted, the mobility saturates to its minimum value of
$ \mu_0 =     \frac{\pi  }{2   k_F^2  } $ \cite{neto1996dynamics}.

Two mobile impurities may be studied by observing that since both are slowly moving, the leading approximation would have only one of them moving and the other fixed so that the applied force would be proportional to the drift velocity of the moving impurity. In general it may be surmised that $ F = const. v_{X} + const. v^2_{X}  $ where the second term is the small correction to the mobility of one of the moving impurity in response to the motion of the second impurity which may be neglected. The mobility of a moving impurity in presence of another
at a distance $ a = \frac{ \xi_0 }{k_F} $ (where $ \xi_0 $ is a tunable dimensionless parameter) is,
\begin{equation}
\begin{aligned}
\mu= &\mu_0 \frac{v_F^2 }{V_0^2 (2V_0^2 \sin^2[\xi_0] -v_F^2)^2} \Big(   (2V_0^2 \sin^2[\xi_0] -v_F^2)^2\\
&+ 4 V_0^2 (V_0 \sin (\xi_0) \cos (\xi_0)+v_F)^2            \Big)
\end{aligned}
\end{equation}
where $\mu\geq \mu_0 \mbox{ } \frac{v_F^2}{V_0^2} $ and the fermions are not mutually interacting.

\begin{figure}[b!]
\begin{center}
\includegraphics[scale=0.5]{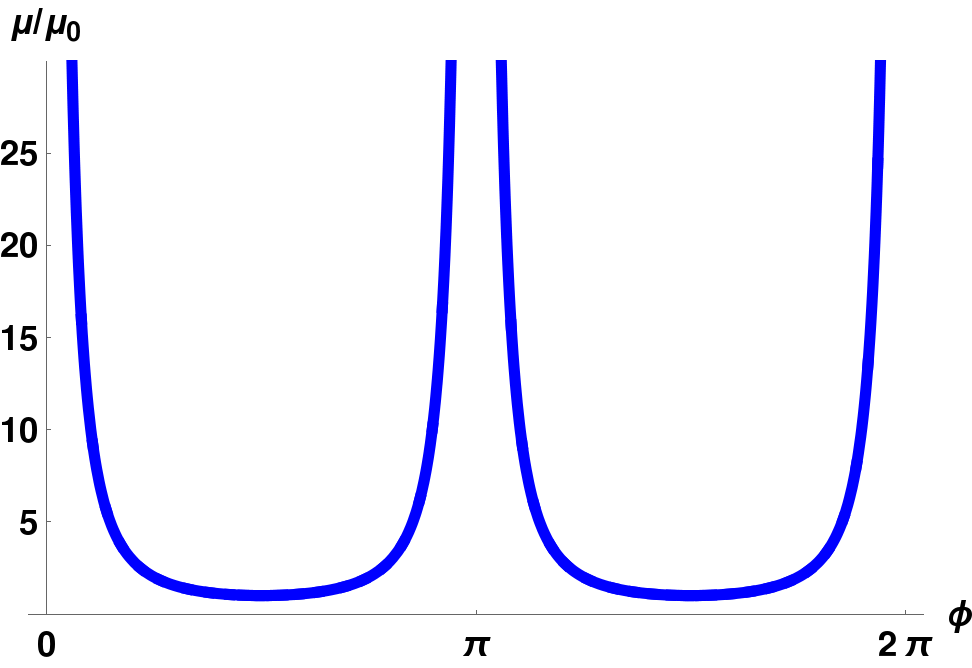}
\end{center}
\caption{Plot of the mobility vs phase $\phi$ for the general case of a finite number of heavy particles.}
\label{phase}
\end{figure}
In presence of a finite number of   identical impurities, the reflection  and transmission amplitudes of fermions may be parametrized as follows: $R = \sin( \theta ) e^{ i \phi } $ and $T=i \cos( \theta ) e^{ i \phi }$ which is consistent with known identities such as $ |T|^2 +|R|^2 = 1 $ and $ RT^{*} = -R^{*}T $.  The mobility then evaluates to a simple expression ($ \mu_0 $ has been defined previously),
\begin{equation}
\mu=\mu_0 \csc^2 \phi
\label{generalMobility}
\end{equation}
The dependence  of the mobility on the phase ($\phi$) is depicted in fig.~ (\ref{phase}). The minimum mobility is $ \mu_0 $  which corresponds to the situation where no tunneling across the impurities is allowed. The maximum mobility is infinite which corresponds to the ballistic motion of the impurity which happens in the trivial situation when the coupling between the impurity and the fermions vanishes. The interesting question is, does it also happen when there are appropriate resonances?  In the two impurity case, an examination of the formula in eq.~(\ref{mobility1}) suggests that ballistic motion of the impurities in response to an applied force may be expected to be seen when $ V_0 =\frac{v_F}{\sqrt{2} \left| \sin (\xi_0)\right| } $, in which case the mobility diverges, but for the most part, the motion of the impurities is heavily damped by the fermions in the background. Also, when $ sin(\xi_0 ) = 0 $, the two-impurity system resembles the single impurity system where there is no possibility of ballistic motion of impurities for non-zero $ V_0 $.  The other extreme is when the strength of the impurities tends to infinity ($V_0 \to \infty$) when the mobility tends to a non-zero minimum value  of $\frac{\pi}{2 k_F^2}$ for a single impurity. Of special interest is the double impurity system where the mobility vanishes when $V_0 \to \infty$, provided $\sin( \xi_0 ) \neq 0$ as may be seen from equation (\ref{mobility1}). This  means that when tunneling through the impurities is forbidden, their mobilities vanish except when the distance between them is an integral multiple of half a Fermi wavelength.

\begin{figure}[h!]
  \centering
  \includegraphics[width=2.7in]{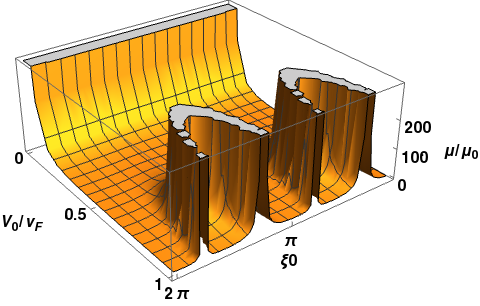}
  \caption{$ \frac{ \mu }{\mu_0 } $ versus $ \frac{V_0}{v_F} $ and $ \xi_0 $ for a two-impurity system}\label{TWOIMP}
\end{figure}
Since the dependence of the phase angle $ \phi $ on $ V_0 $ and $ \xi_0 $ in the two-impurity case is complicated, it is better to visualize the dependence of the mobility on these variables more directly through the following 3D plot in fig.~ \ref{TWOIMP}. The explicit dependence of the phase angle $ \phi $ on $ V_0 $ and $ \xi_0 $ for a two impurity system may be expressed as follows.
\begin{equation*}
\sin ^2(\phi )=\frac{(2V_0^2 \sin^2(\xi_0) -v_F^2)^2 }{\left(v_F^4+4V_0^2(v_F \cos{(\xi_0)}+V_0 \sin{(\xi_0)})^2\right)}
\end{equation*}

\section{ Mobility of a single and two-impurity system in a Luttinger liquid}

When mutual interactions between the fermions are included, it is well-known that the low temperature properties of these systems are qualitatively different from those of free fermions. They form what is known as a Luttinger liquid.  In earlier works \cite{das2016quantum,das2017one}, the present authors have shown that the most singular part of the RPA Green function of Luttinger liquids in presence of barriers or wells is a discontinuous function of the height of the barriers for small barrier heights. Thus the limit of small barriers (large tunneling amplitudes) may not be usefully studied by the  traditional approaches that invoke a perturbation theory  around the homogeneous Luttinger liquid starting point. In fact the present approach which is based on a non-standard harmonic analysis of the fast part of the density fluctuations is uniquely suited to study impurity systems as it allows for an analytical interpolation between the weak link and the weak barrier extreme limits unlike the traditional approaches that fall well short of providing explicit expressions for the exponents associated with the mobility in the general cases and instead rely on tentative renormalization flow analyses.

 It suffices to state that the analytical expressions for the Green functions for Luttinger liquids in presence of barriers and wells derived in a recent work \cite{das2016quantum} are borrowed and used as input to compute the mobility of one and two impurities using the algorithm outlined in Sec.II of the present work. This leads to the following formula for the external force acting on the impurity $ F_X $ in terms of the drift velocity $ v_X $.
\begin{equation*}
F_X \sim \mu_{*}^{-1} \mbox{  }\omega^{\alpha } \mbox{   }v_X
\end{equation*}
where $ \mu_{*} $ is the mobility of the corresponding system with no mutual interactions between fermions and,
\begin{equation*}
\alpha = \text{Min}(\alpha_1 , \alpha_2 )
\end{equation*}
The explicit values of $\alpha_1$ and $\alpha_2$  both for single and two impurities are given in Appendix I. Here $ \alpha_1 $ is the dynamical density of states (DDOS) exponent at the origin when the two (spatial) points of the two-point function are assumed to merge with the origin from the same side while $\alpha_2$ is that from the opposite sides. Both these appear in the analysis since in the defining equation for the Green function viz. eq.~(\ref{Keldysh}), even if $ x = x' $, the heavy particle can be found on either side of $ x = x' $.

\begin{figure}[h!]
  \centering
  \includegraphics[width=2.3 in]{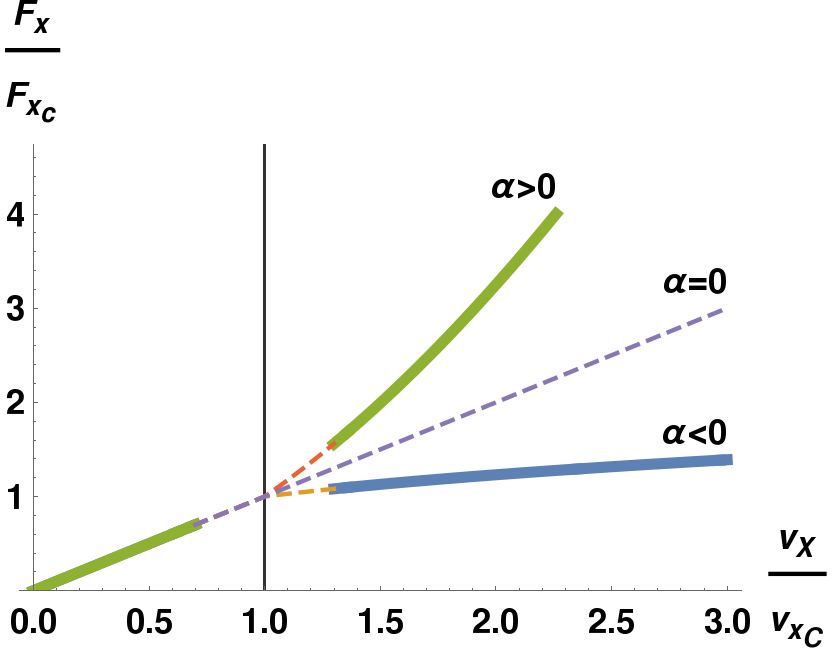}
  \caption{Plot of $\frac{F_{X}}{F_{X_c}}$ vs $\frac{v_{X}}{v_{X_c}}$ where $v_{X_c}=\frac{T}{k_F}$ is constant for a given temperature. The dotted lines signify regions that interpolate between regimes that are easily amenable to analytical approaches.}
  \label{FXvsVX}
\end{figure}

As pointed out earlier, $ \omega $ is the dominant energy scale that corresponds to temperature as long as $ \frac{ k_F v_X }{ T } \ll 1 $ but $ \omega $ would correspond to the energy scale set by the drift velocity viz. $  v_X k_F $ if $ \frac{ k_F v_X }{ T } \gg 1 $. Roughly speaking it should not matter whether the drift velocity or the externally applied force is used to pin down the second energy scale (other than temperature), these two notions are interchangeable as long as $ |\alpha | \ll 1 $. The present study is limited to regions where this condition is obeyed.
Thus the force acting on the heavy particle is explicitly expressed as follows.
\begin{equation}
\begin{aligned}
F_X=\mu_{*}^{-1}\left[Max\left(\frac{k_F v_X}{v_F \Lambda},\frac{T}{v_F \Lambda}\right)\right]^{Min(\alpha_1,\alpha_2)} \mbox{ } v_X
\label{choice}
\end{aligned}
\end{equation}
Here $v_{X}$ is the drift velocity of the heavy particle(s) when acted upon by a force $ F_X $. The cross-over speed which provides a scale that separates the regime of linear dependence on the applied force and the nonlinear regime is clearly,  $v_{X_c}=\frac{T}{k_F}$. The force on the particle moving with this cross-over speed is the cross-over force
\begin{equation*}
F_{X_c}=\mu_{*}^{-1}\left[\frac{T}{v_F \Lambda}\right]^{Min(\alpha_1,\alpha_2)} \mbox{ }v_{X_c} 
\end{equation*}
where $\Lambda v_F$ is the band-width mentioned earlier. fig.~ (\ref{FXvsVX}) shows the variation of the force on the heavy particle as a function of the drift velocity (rescaled appropriately to make them both, dimensionless).
From fig.~ (\ref{FXvsVX}), it is clear that when $v_X \ll v_{X_c}=\frac{T}{k_F}$, the force varies linearly with drift velocity, the ratio being the inverse of mobility. On the other hand when $v_X \gg v_{X_c}$, the force varies non-linearly with the drift velocity and the curvature is decided by the sign of the exponent $\alpha$.

\begin{figure}[h!]
  \centering
  \includegraphics[width=2.3 in]{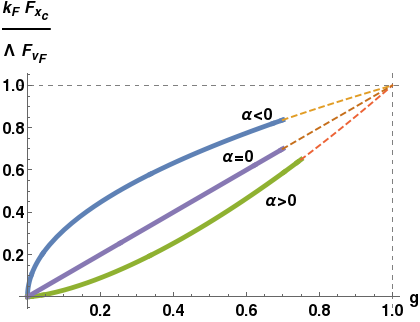}
  \caption{Variation of the cross-over force scale on temperature for different choices of the sign of $\alpha$.}
  \label{FXCvsg}
\end{figure}

The dependence of the cross-over scales themselves on temperature may also be studied. Consider the dimensionless quantities $ \frac{ k_F v_{X_c} }{ \Lambda v_F } $ and $ \frac{ k_F F_{X_c}  }{ \Lambda F_{v_F} } $ where $ F_{v_F} $ is the hypothetical force extrapolated from the above formulas which would naively correspond to the force acting on a heavy particle whose drift velocity is the Fermi velocity $ v_F $ (there is no valid physics here - this is just a scale to render dimensional quantities, dimensionless). It is pertinent to examine the dependence of these quantities on the dimensionless temperature $ g = \frac{T}{\Lambda v_F} $. From the above formulas it is clear that,
\begin{equation}
\frac{ k_F F_{X_c}  }{ \Lambda F_{v_F} } = g^{ 1 + \alpha }; \mbox{   } \mbox{   }\mbox{   } \mbox{   }
 \frac{ k_F v_{X_c} }{ \Lambda v_F } = g
\end{equation}
These formulas may also be  described diagrammatically as in fig.~(\ref{FXCvsg}).
An examination of fig.~ ({\ref{mobilityvsg}) shows that for negative values of $\alpha$, the mobility decreases from a maximum value with an increase in temperature while for positive values of $\alpha$, it  increases from a minimum value with an increase in temperature, while both of them tend to converge to the value of the non-interacting case.
\begin{figure}[h!]
  \centering
  \includegraphics[width=2.3 in]{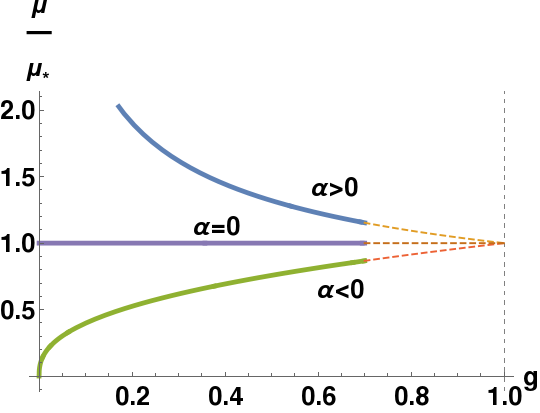}
  \caption{Ratio of mobility in the linear regime in presence of interactions to that without interactions vs $g$ (temperature).}
  \label{mobilityvsg}
\end{figure}
As mentioned earlier, the force acting on the heavy particle is expressed as a power-law in terms of the energy scale `$\omega$' which is expressed in units of the bandwidth $\Lambda v_F$ such that $\frac{\omega}{\Lambda v_F}$ is dimensionless and is always less than unity. The expression for force obtained from the Green function is originally a linear combination of terms with different powers of  $\frac{\omega}{\Lambda v_F}$ and thus the dominant term is the one with the smallest exponent, which explains the choice of $\alpha$ as the minimum of $\alpha_1$ and $\alpha_2$ in equation (\ref{choice}).
\begin{figure}[b!]
  \centering
  \includegraphics[width=3 in]{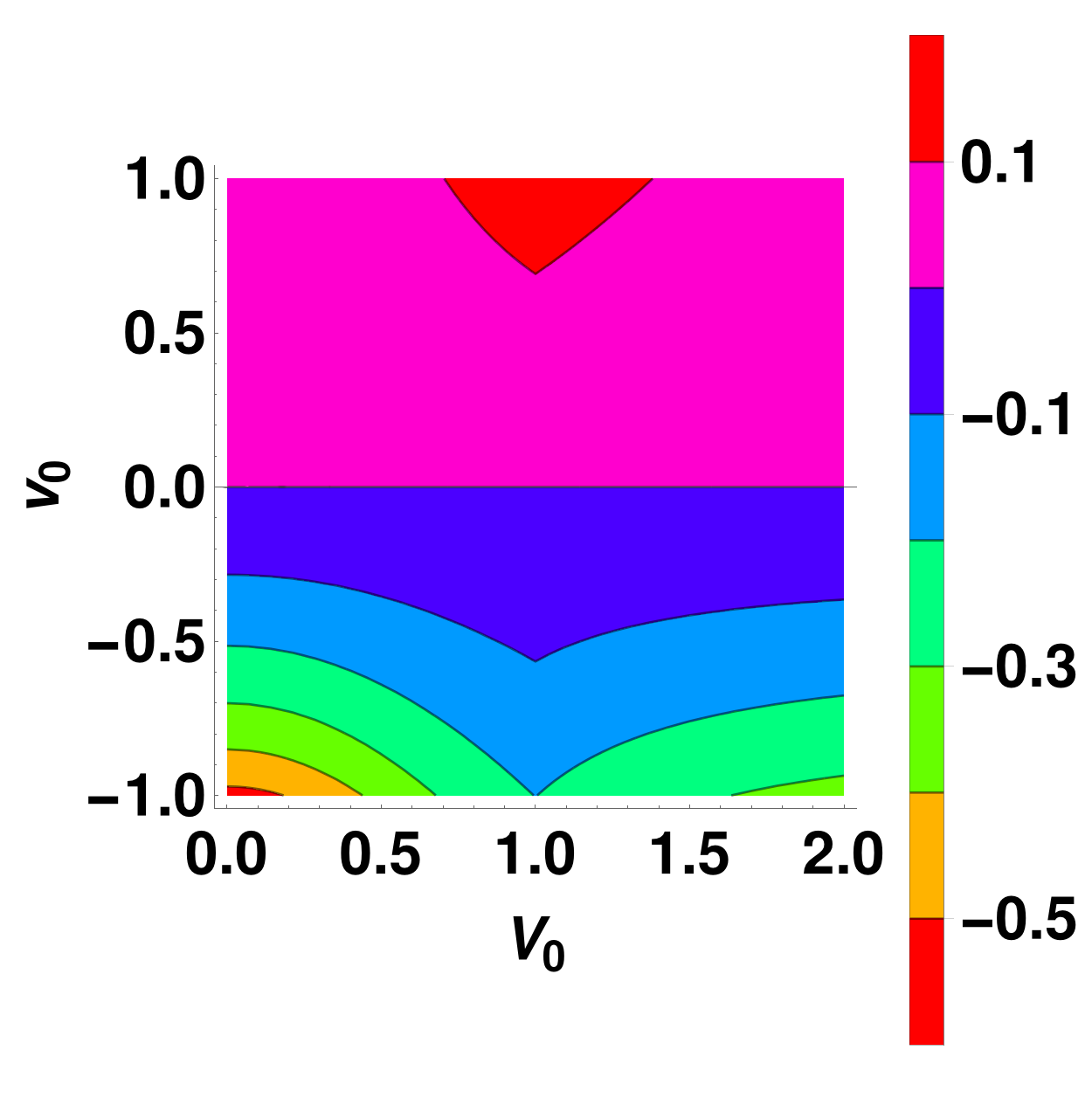}\\(a)\\
  \includegraphics[width=3 in]{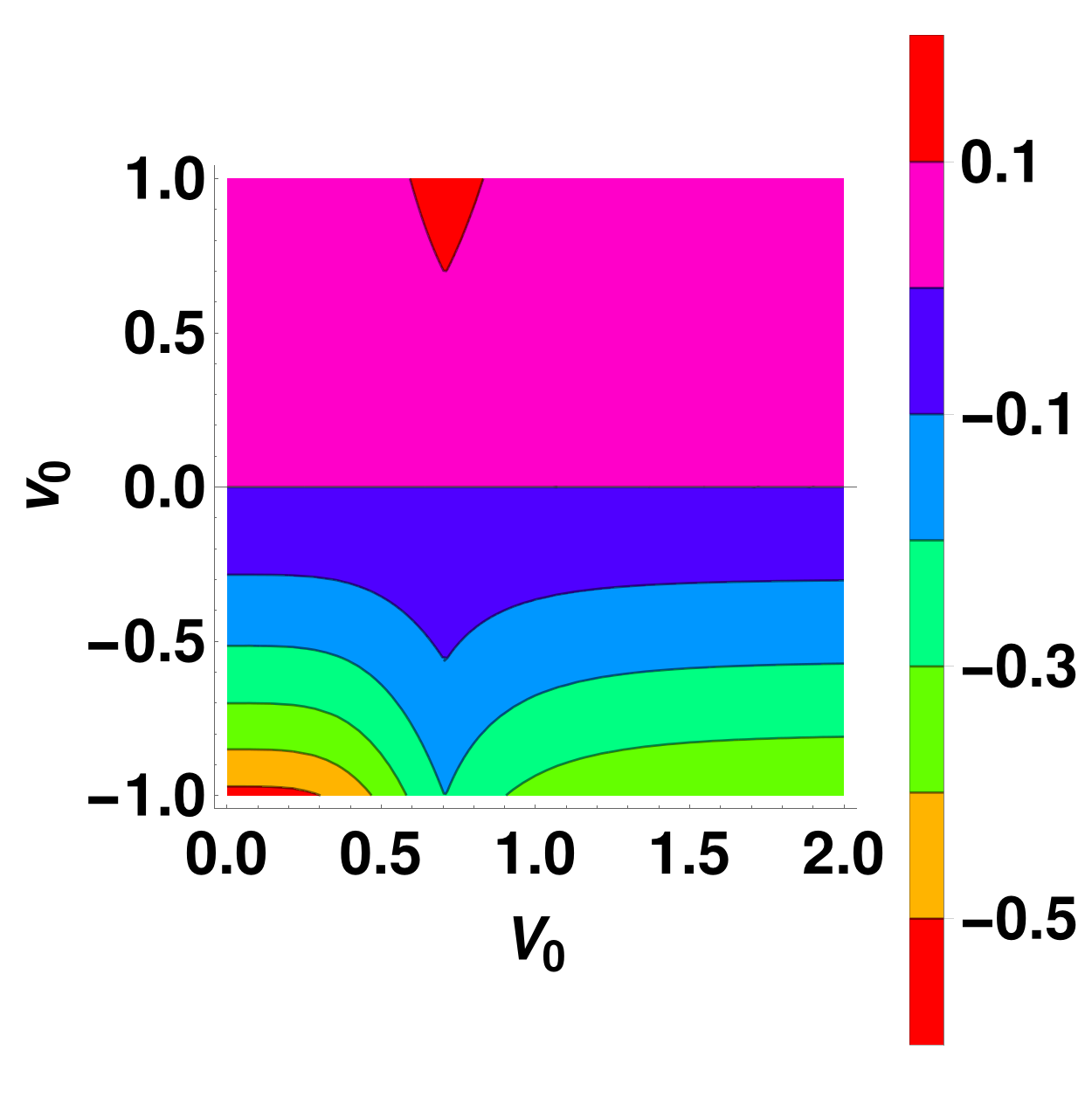}\\
(b)\\
  \caption{Plots of the exponent $\alpha$ for (a) Single impurity system and (b) Two-impurity system as a function of the impurity strenth $V_0$ and  the strength of the mutual interactions $v_0$ (setting $v_F=1$ and $\xi_0=\pi/2 + n \pi $ for two impurity case where $n$ is an integer).}
  \label{singledouble}
\end{figure}
\begin{figure}[t!]
  \centering
  \includegraphics[width=3 in]{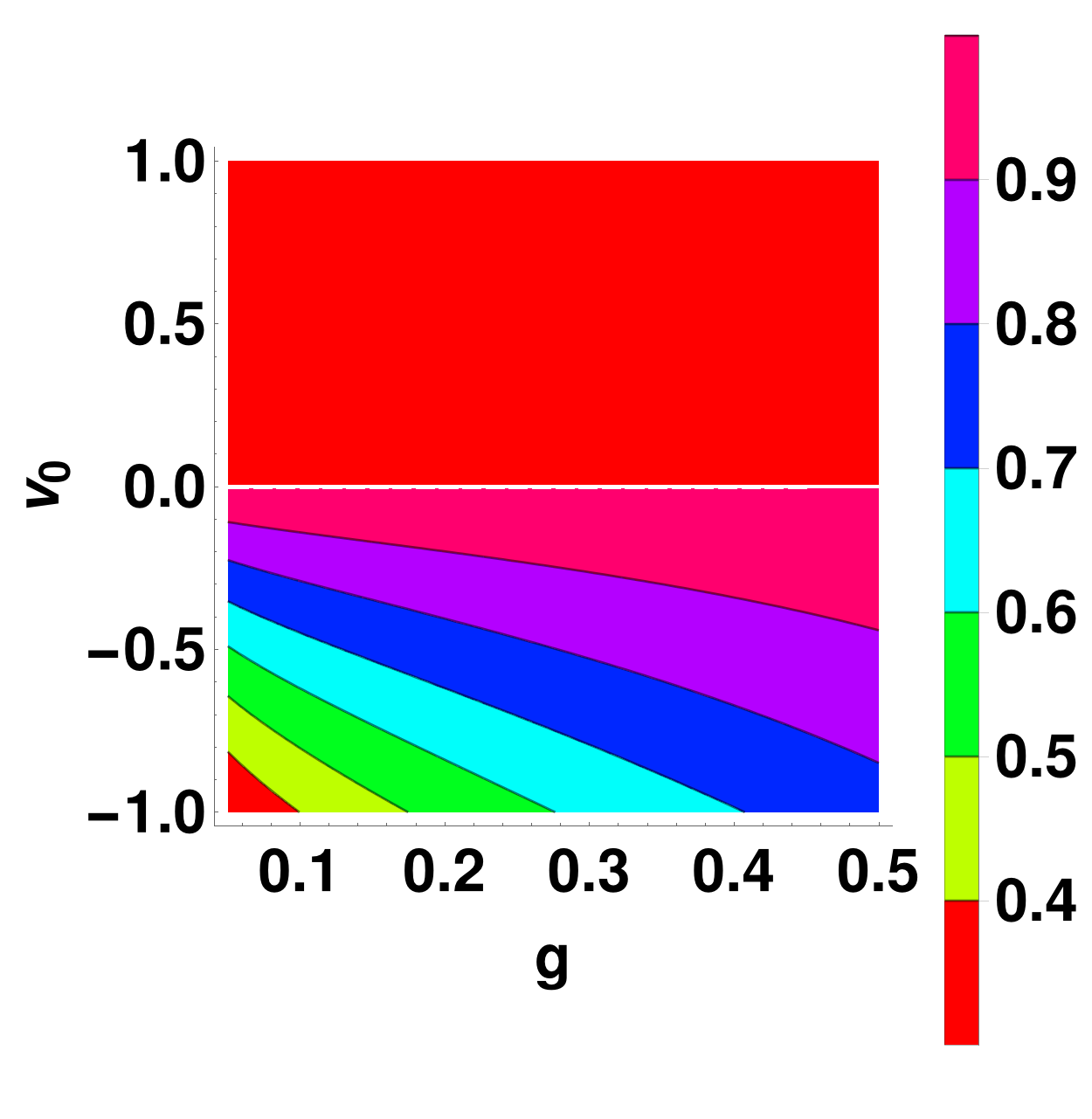}
  \caption{Ratio of mobility in the linear regime in presence of interactions to that without interactions vs $g$ (temperature) and  the strength of the mutual interactions $v_0$ when the strength of impurities diverge (no-tunneling case).}
  \label{minus}
\end{figure}

It is important to analyze the exponent $\alpha$ since earlier plots show that the sign of $\alpha$ is quite important in determining the qualitative behavior of the applied force versus terminal velocity. In general, $\alpha$ can be positive or negative or even vanish altogether while its absolute value is sufficiently less than unity. From the plots in fig.~ (\ref{singledouble}) it is observed that both for single and double impurity, $\alpha$ tends to take positive values for repulsive interactions and thus the mobility, as observed from fig.~ (\ref{mobilityvsg}), decreases with an increase in temperature which is  consistent with the literature \cite{ neto1994mobility, neto1996dynamics}. On the other hand for attractive interactions, $\alpha$ takes negative values which indicates an increase in mobility with an increase in temperature. The two-impurity system shows some interesting physics in the behavior of the exponents as well. The plot in fig.~ (\ref{singledouble}b) is given for $\xi_0=\pi/2$ but it is observed that the exact same plot is obtained for $\xi=\pi/2 + n \pi$ where n is an integer. This indicates that the mobility oscillates as a function of $\xi_0$ with a period of $\pi$.

Finally the mobility is studied  when tunneling through the impurities is forbidden ($V_0 \to \infty$). In this case the exponent $\alpha$ vanishes for repulsive interactions and becomes equal to $(v_h-v_F)/v_F$ for attractive interactions both for the double and the single impurity. The double impurity is of lesser importance here because, as already discussed earlier, the mobility vanishes for double impurity in this situation except when $\xi_0=n \pi$. The variation of mobility is shown in fig.~ (\ref{minus}) as a function of the forward scattering strength $v_0$ and temperature $g$ ($= \frac{kT}{\Lambda v_F}$).

Generally, fermions that are mutually attracting tend to mitigate the effect of an impurity \cite{kane1992transport} (``heal the chain"). A weak impurity in turn implies a tendency toward ballistic mobility or at least increased mobility. At higher temperatures, attractive fermions become better at mitigating the effect of the impurity hence the mobility increases when temperature increases.

Conversely for fermions that are mutually repelling, there is a tendency to aggravate the effect of the impurity \cite{kane1992transport} (``cut the chain"). However, when the impurity strength is already strong enough to prevent tunneling through the impurity, the mutual repulsion of the fermions do not do anything to the mobility as it is already the minimum value it can be. Hence in this situation the mobility is independent of temperature.
 When tunneling is allowed in the repulsive case, mobility decreases with increasing temperature as it would have done had the barrier strength increased instead.

\section{Comparison with existing studies}
 The highly cited work on the subject by Castro-Neto et al. \cite{neto1996dynamics} considers a.c. mobility. When the applied force is a.c., the terminal velocity is also a.c. and proportional to the applied force right down to absolute-zero temperature. This is not the case in the present work where we consider a d.c. applied force. At temperatures small compared to the Fermi energy when fermion-fermion interactions are ignored, it is well-known that the mobility of a heavy particle is temperature independent. This is clearly stated in \cite{neto1996dynamics}, and their $\mu_0$ is the same as ours - indeed we simply borrowed this well-known result.  At very low temperatures the main prediction of \cite{neto1996dynamics} is that the a.c. mobility (d.c. limit of a.c.  mobility is not the same as d.c. mobility: this is clearly stated in \cite{neto1996dynamics}. We have calculated d.c. mobility and not the d.c. limit of a.c. mobility) diverges as the 4-th power of temperature whereas at high temperatures it is approximately independent of temperature as long as the heavy particle is impenetrable by the mutually repelling fermions.
In order to study this limit as best as we can using our manifestly d.c. formulas, we first observe that the terminal velocity $v_X$ is related to applied d.c. force $F_X$ at very low temperatures in the following nonlinear way:
\begin{equation*}
 F_X = \mu^{-1}_{*} v_X^{1+c}
\end{equation*} where   $c=(v_h^2-v_F^2)/(2 V_0^2)$ for $v_h > v_F$ (repulsion between fermions) and $c=(v_h-v_F)/v_F$ for $v_h < v_F$ (attraction between fermions), both for an impenetrable impurity ($V_0 \to \infty$).
Now the differential mobility is 
\begin{equation*}
\mu = \frac{dv_X}{dF_X} = (1+c)^{-1} \mu_{*} v_X^{-c}
\end{equation*}
 The linear mobility is defined as the $v_X \to  0$ limit  of the differential mobility.
\begin{equation*}
\mu_{diff}= \lim_{ v_X \to 0 } (1+c)^{-1} \mu_{*} v_X^{-c} = \infty
\hspace{0.1cm} \text{since   } c > 0\mbox{ } (repulsion) 
\end{equation*}
\\
This is consistent with \cite{neto1996dynamics} which says that at very low temperatures the mobility of an impenetrable heavy particle with mutually repelling fermions diverges (motion is ballistic). Conversely at high temperatures, both \cite{neto1996dynamics} and our paper predicts a roughly temperature independent linear mobility as long as the impurity is impenetrable by the mutually repelling fermions. The particular result of \cite{neto1996dynamics}  namely the $ T^{-4} $ law is derived by them by treating the time-dependent spatially inhomogeneous impurity potential as a small perturbation around the homogeneous Luttinger liquid background. This is because their RG equations show that at low temperature the impurity behaves effectively as if it were much lighter and much more penetrable.   Since our formalism is identical to theirs for the homogeneous system, discussing this $ T^{-4} $ law would be a simple duplication of their analysis. Our results are only valid for a fully d.c. externally applied force and hence this is qualitatively different from the situation they consider in the latter half of their paper. Even so, our results also confirm their conclusions namely that the impurity tends to be much more mobile at low temperatures when fermions are mutually repelling than when they are non-interacting. \\

\section{Conclusions}

In this work, the Green function of slowly moving impurities in a Luttinger liquid is obtained using a combination of perturbative approach and the non-chiral bosonization technique. The force acting on the heavy particle is calculated as a function of the drift velocity for the non-interacting case and the expression for mobility is calculated. Both the linear and non-linear dependence of the force on the drift velocity has been analytically obtained for systems with forward scattering interactions between the fermions with one or two mobile impurities.
Peculiar resonances that are seen in the two-impurity system have been mapped out. The unique feature of this work is the analytical closed expressions for the exponents in terms of the coupling strengths in the problem that interpolate between the ballistic regime on the one hand and the no-tunneling regime on the other.

\section*{Funding}
   A part of this work was done with financial support from Department of Science and Technology, Govt. of India DST/SERC: SR/S2/CMP/46 2009.\\
\section*{Appendix I: Exponents $\alpha_1$ and $\alpha_2$}
For one impurity,
\begin{equation*}
\begin{aligned}
\alpha_1=&\frac{(v_h-v_F)(2V_0^2+v_F(v_h-v_F))}{2v_F(V_0^2+v_hv_F)};
\alpha_2=\frac{v_h^2-v_F^2}{2(V_0^2+v_hv_F)}\\
\end{aligned}
\end{equation*}
where $v_h=v_F\sqrt{1+\frac{2v_0}{\pi v_F}}$, $V_0$ is the strength of the impurity and $v_0$ is the strength of forward scattering interactions such that $v_0<0$ for attractive interactions and $v_0>0$ for repulsive ones.

For two identical impurities,
\begin{equation*}
\begin{aligned}
\alpha_1=&\frac{v_h-v_F}{v_F}+\frac{v_F^2(v_F^2-v_h^2)}{2(4V_0^2(V_0 \sin(\xi_0)+v_F \cos (\xi_0))^2+v_F^3v_h)}\\
\alpha_2=&\frac{v_F^2(v_h^2-v_F^2)}{2(4V_0^2(V_0 \sin(\xi_0)+v_F \cos (\xi_0))^2+v_F^3v_h)}\\
\end{aligned}
\end{equation*}
with $\xi_0=k_F a$ where $a$ is the distance between the two impurities and $k_F$ is the fermi momentum.


\bibliographystyle{apsrev4-1}
\bibliography{ref}
\normalsize

\end{document}